# Interaction Ruling Animal Collective Behaviour Depends on Topological rather than Metric Distance: Evidence from a Field Study

M. Ballerini[1,2], N. Cabibbo[3,4], R. Candelier[3]†, A. Cavagna[1,5]★, E. Cisbani[2], I. Giardina[1,5], V. Lecomte[6]†, A. Orlandi[1], G. Parisi[1,3,4], A. Procaccini[1,3], M. Viale[3]† & V. Zdravkovic[1]

[1] Centre for Statistical Mechanics and Complexity (SMC), CNR-INFM, Dipartimento di Fisica, Universita' di Roma 'La Sapienza', Piazzale Aldo Moro 2, 00185 Roma, Italy

[2] Istituto Superiore di Sanita', viale Regina Elena 299, 00161 Roma, Italy

[3] Dipartimento di Fisica, Universita' di Roma 'La Sapienza', Piazzale Aldo Moro 2, 00185 Roma, Italy

[4] Sezione INFN, Universita' di Roma `La Sapienza', Piazzale Aldo Moro 2, 00185 Roma, Italy

[5] Istituto dei Sistemi Complessi (ISC), CNR, via dei Taurini 19, 00185 Roma, Italy

[6] Laboratoire Matière et Systèmes Complexes, (CNRS UMR 7057), Université Paris VII, 10 rue Alice Domon et Léonie Duquet, 75205 Paris Cedex 13, France

★ To whom correspondence should be addressed. E-mail: andrea.cavagna@roma1.infn.it

† Present addresses: R.C.: GIT / SPEC / DRECAM, Bat. 772, Orme des Merisiers, CEA Saclay, 91191 Gif sur Yvette, France; M.V.: Dipartimento di Fisica, Universita' di Roma 3, via della Vasca Navale 84, 00146 Roma, Italy; V.L.: DPMC, Université de Genève, 24 Quai Ernest Ansermet, 1211 Genève, Suisse.

2**Numerical models indicate that collective animal behaviour may emerge from simple local rules of interaction among the individuals. However, very little is known about the nature of such interaction, so that models and theories mostly rely on aprioristic assumptions. By reconstructing the three-dimensional position of individual birds in airborne flocks of few thousands members, we prove that the interaction does not depend on the metric distance, as most current models and theories assume, but rather on the topological distance. In fact, we discover that each bird interacts on average with a fixed number of neighbours (six-seven), rather than with all neighbours within a fixed metric distance. We argue that a topological interaction is indispensable to maintain flock's cohesion against the large density changes caused by external perturbations, typically predation. We support this hypothesis by numerical simulations, showing that a topological interaction grants significantly higher cohesion of the aggregation compared to a standard metric one.**

**Numerical models indicate that collective animal behaviour may emerge from simple local rules of interaction among the individuals. However, very little is known about the nature of such interaction, so that models and theories mostly rely on aprioristic assumptions. By reconstructing the three-dimensional position of individual birds in airborne flocks of few thousands members, we prove that the interaction does not depend on the metric distance, as most current models and theories assume, but rather on the topological distance. In fact, we discover that each bird interacts on average with a fixed number of neighbours (six-seven), rather than with all neighbours within a fixed metric distance. We argue that a topological interaction is indispensable to maintain flock's cohesion against the large density changes caused by external perturbations, typically predation. We support this hypothesis by numerical simulations, showing that a topological interaction grants significantly higher cohesion of the aggregation compared to a standard metric one.**

**Introduction**

Collective behaviour of large aggregations of animals is a truly arresting natural phenomenon (1). Particularly interesting is the case when aggregations self-organize into complex patterns with no need of an external stimulus. Prominent examples of such behaviour are bird flocks (2), fish schools (3) and mammal herds (4). Apart from its obvious relevance in ethology and evolutionary biology, collective behaviour is a key concept in many other fields of science, including physics (5), control theory (6), mobile robotics (7), economics (8), and social sciences (9).

How does collective behaviour emerge? Numerical models of self-organized motion, inspired both by biology (10,11,12,13,14,15,16), and by physics (17,18,19), support the idea that simple rules of interaction among the individuals are sufficient to produce collective behaviour. Unfortunately, we have very scarce empirical information about the precise nature of such rules. The main theoretical assumptions (attraction among the individuals, short range repulsion, and alignment of the velocities) are reasonable, but generic, and there are as many different models as different ways to implement these assumptions. Without decisive experimental feedback it is difficult to select what is the 'right' model, and therefore to understand what are the underlying fundamental rules of animal collective behaviour.

The main goal of the interaction among individuals is to maintain cohesion of the aggregation. This is a very strong biological requirement, shaped by the evolutionary pressure for survivor: stragglers and small groups are significantly more prone to predation than animals belonging to large and highly cohesive aggregations (3,20). Consider a flock of starlings under attack by a peregrine falcon: the flock contracts, expands and even splits, continuously changing its density and structure. Yet, no bird remains isolated, and soon the flock reforms as whole. The question we want to answer is: what kind of interaction maintains cohesion in such a *robust* way?

Dramatic density changes as the ones observed in animal aggregations are not at all common in physical systems: the density of an aggregate of particles is changed by changing the external constraints, as the volume of the box containing a gas. When, on the other hand, particles condensate in open space, the density is determined mainly by the dependence of the inter-particle force on the distance, dependence that is a characteristic of the material and is therefore fixed. If the aggregation is subject to a strong external perturbation, density does not vary, but cohesion breaks down. Flocks and schools, however, live in an open space, and yet undergo enormous density changes without losing cohesion, in a very flexible way. This seems a challenging problem when tackled within the framework of standard physical interactions.

To grant cohesion, models make the sound assumption that individuals align and attract to each other, and that such interaction decays with increasing distance between individuals. The vast majority of models, both developed by physicists and biologists, adopt a definition of 'distance' that is the same as in physics, i.e. *metric* distance. In a metric context, two birds 5 meters apart attract each other less than two birds 1 meter apart. Animals can estimate metric distance in various ways, including stereovision, retinal image size, and optic flow (21). Thus, a metric interaction seems natural. However, physical systems warn us that an interaction based on metric distance may be unable to reproduce the density changes typical of animal aggregations. An alternative is *topological* distance. In economics, for example, the relevant quantity is not how many kilometres separate two countries (metric distance), but rather the *number* of intermediate countries between them (topological distance) (22). If the interaction depends on the topological distance, each individual interacts with a fixed number of neighbours, *irrespective of their metric distance*. The crucial difference between metric and topological interaction really kicks in when the density varies: in the topological case, two birds 5 meters apart in a *sparse* flock attract each other as much as two birds 1 meter apart in a *denser* flock, provided that the number of individuals between the two birds is the same. The strength of the interaction remains the same at different densities. This seems more suitable to keep cohesion in the face of strong density fluctuations. By means of empirical observations we show that the topological paradigm is in fact the correct one.

Structure is the foremost effect of interaction, and, conversely, interaction is ciphered in the inter-individual spatial structure. Hence, in order to learn something about the interaction ruling collective behaviour it is necessary to analyze the structural organization of individuals within the aggregation. To do this, however, it is essential to have data on the 3D positions of individuals in *large* aggregations: collective behaviour is a qualitatively different phenomenon, with emerging complex patterns, only when the number of individuals is big; moreover, in small aggregations the surface-to-volume ratio is large, and the bias introduced by the border is inevitably very strong (see Methods). Unfortunately, gathering quantitative 3D data on even moderately large groups of animals is very difficult. Most empirical studies have two major limitations: a small number of individuals (few tens), and loose group arrangements (23,24), at variance with the huge, highly cohesive natural aggregations.

**Results**

Thanks to novel stereometric and computer vision techniques, we measured 3D individual birds positions in compact flocks of up to 2600 starlings (*Sturnus vulgaris*) in the field. This is an advance of almost two orders of magnitudes compared to former experiments. A typical flock and its 3D reconstruction are shown in Fig.1 (see also Figs.S3 in Supplementary Material [SM]). Starlings' aerial display provides a paradigmatic case of collective behaviour (25). These birds gather in the evening over the roost and form sharp-bordered, strongly cohesive flocks, ranging from few hundreds to tens of thousands birds (see SM Figs.S1-S2). We reconstructed and analyzed ten flocking events recorded at the roosting site of Termini railway station (Rome, Italy) between Dec. 2005 and Feb. 2006. Each event is defined by a series of up to 80 stereo photographs, shot at 10 frames-per-second. Different events correspond to different flocking flight sequences, recorded on separate sessions. Observations were done at dusk. Details on the experiment and on the reconstruction algorithms can be found in the Materials and Methods section.

The clearest characterization of the structure of birds within a flock is given by the spatial distribution of the nearest neighbours. Given a reference bird, we measure the angular orientation of its nearest neighbour with respect to the flock's direction of motion, i.e. the neighbour's bearing and elevation. We repeat this by taking all individuals within a flock as reference bird, and in this way we map the average spatial position of nearest neighbours (see caption of Fig.2). This map (Fig.2a) shows a striking lack of nearest neighbours along the direction of motion. The structure of individuals is therefore strongly anisotropic. The possible reasons for this anisotropy, probably related to the visual apparatus of birds, are discussed in SM. The crucial point, however, is that this anisotropy is the effect of the interaction among individuals. To support this claim, we compute the distribution of neighbours very far apart from the reference bird (Fig.2b). This distribution is uniform, as for a completely isotropic, non-interacting aggregation of points. This is a direct empirical indication that interaction decays with the distance, and it demonstrates that we can use the anisotropy to get information about the interaction (see also SM on this point).

To quantify the decay of the anisotropy we define a function $\gamma(n)$ that measures to what extent the spatial distribution of the $n^{\text{th}}$-nearest neighbour around a reference bird is anisotropic (see caption of Fig.3). The value of $\gamma$ for an isotropic, non-interacting aggregation is 1/3. A value larger than 1/3 indicates that the interaction among the birds makes the structure anisotropic. In Fig.3a we show that $\gamma(n)$ decays gradually to 1/3 when $n$ increases. Hence, for each flock we can define an interaction range $n_c$, given by the value of $n$ where $\gamma$ becomes 1/3. By definition, birds farther than the $n_c^{\text{th}}$ nearest neighbour are isotropically distributed around the reference bird, and do not interact with it. We note that Fig.3a is the first empirical determination of how the interaction decays in a real instance of collective behaviour.

The $n^{\text{th}}$ nearest neighbour of a given bird is characterized not only by its integer label $n$, but also by its actual distance in meters $r$ from the reference bird. For example, in flock 32-06 (Fig.3b) the $6^{\text{th}}$ nearest neighbour of a bird is found, on average, at 1.25m from it. Clearly, the relation between $n$ and $r$ depends on the specific density of the flock. While $n$ measures the *topological* distance from a reference bird, $r$ measures the *metric* distance. In addition to the topological interaction range in unit of birds, $n_c$, we can therefore introduce a metric range, in unit of meters, $r_c$. Going back to flock 32-06, we have $n_c=6$, and $r_c=1.25$m (Fig.3a).

The flocks we analyzed differ a lot one from the other in density. This implies that the topological and metric ranges, $n_c$ and $r_c$, cannot be *both* constant from flock to flock. To elucidate this crucial point, let us consider two flocks with different densities. If the interaction depends on the *metric* distance, then the range in meters $r_c$ is the same in the two flocks, while the number of individuals $n_c$ within this range is large in the denser flock, and small in the sparser one. Conversely, if the interaction depends on the *topological* distance, the range in units of birds $n_c$ is constant in the two flocks, while the distance $r_c$ of the $n_c^{th}$ nearest neighbour is small in the denser flock, and large in the sparser one. The difference between topological and metric hypothesis is stark: in the topological scenario the number of interacting individuals is *fixed*. On the opposite, in the metric scenario, such number *varies* with density; for example, within the same metric range there are 10 birds in our densest flock, and only 1 bird in the sparsest one. Topological and metric ranges therefore are *not* interchangeable characterizations of the interaction. Therefore, to understand whether it is the metric or the topological distance that matters we must measure how $r_c$ and $n_c$ depend on the flocks' density.

To cast in a quantitative way the two opposite scenarios, we note that the average distance $r$ of the $n^{th}$ nearest neighbour grows with $n$ according to the relation $r \sim r_1 n^{1/3}$ (see Fig.3b). In this equation $r_1$ is the average nearest-neighbours distance, which is a direct measure of sparseness (the inverse of density); $r_1$ varies from 0.68m in the densest flock, to 1.51m in the sparsest one (see table S1 in SM). The equation above simply means that the number $n$ of individuals within a sphere of radius $r$ is proportional to $r^3$. The two ranges are linked by the same relation, $r_c \sim r_1 n_c^{1/3}$. In a metric scenario, $r_c$ is a constant, and thus $n_c^{-1/3} \sim r_1$. Conversely, in the topological scenario $n_c$ is a constant, and thus $r_c \sim r_1$. We have measured $n_c$ and $r_c$ in each flock, and have studied how these two quantities depend on the flocks' sparseness $r_1$. The experimental evidence clearly supports the topological scenario: there is no significant correlation between $n_c^{-1/3}$ and $r_1$ (Pearson's correlation test: $n=10$, $R^2=0.00021$, $P=0.97$), whereas a clear linear correlation exists between $r_c$ and $r_1$ ($n=10$, $R^2=0.78$, $P=0.00072$) (Fig.3c,3d). The topological range is therefore approximately constant from flock to flock. On average, we find $n_c=6.5\pm0.9$ SE.

We therefore showed that the interaction depends on the topological distance, rather than the metric one. The interaction between two birds 1m apart in flock *A,* is as strong as that between two birds 5m apart in flock *B*, provided that *A* is denser than *B* and that the topological distance $n$ is the same. Our empirical result contrasts with the assumption of most models and theories. Even though some models introduce a cut-off, or numerical preference, in the number of interacting neighbours (so that this number is fixed), they still 'weight' these neighbours metrically (26). We must stress that this is *not* what we find here. It is the very shape of the interaction that depends on the topological distance, not simply the cut-off, or the range (Fig.3a). Our result also rules out for starling flocks the hypothesis that the anisotropic structure is a consequence of the bird's effort to take advantage of the wakes of its neighbours (27), since such aerodynamic advantage would decay within a well-defined metric length-scale (see SM for a discussion of this point). In fact, we believe that the only mechanism compatible with our result is vision.

**Discussion**

Why 6-7 neighbours? This range is significantly smaller than the number of visually unobstructed neighbours around each bird. We conclude that this specific value of $n_c$ must derive from the cortical elaboration of the visual input, rather than from a limitation of the input itself. In order to keep under control a fixed number of neighbours, irrespective of their metric distance, it is necessary for the individuals to have some pre-numeric ability, or, more precisely, an object-tracking, or 'subitizing', ability (28). This capability decays beyond a certain number, and such perceptual limit defines the range of interaction. Laboratory experiments show that trained pigeons can discriminate sets of different numerosities provided that these sets have less than 7 objects (29). In our field study we find a range of 6-7 neighbours. Such a striking agreement suggests that the same tracking ability at the basis of numerical discrimination may be used for interacting with a fixed number of neighbours, and then be an essential ingredient of collective animal behaviour. The existence of a perceptual limit in numerosity is also found in 2D experiments on shoaling fish, and it is estimated around 3-5 individuals (30). An alternative interpretation of the interaction range we find is that the specific value of $n_c$ may be functional to optimize anti-predatory response: if each individual interacts with too few neighbours, information is non-noisy, but too short-ranged; conversely, if the interaction involves too many neighbours, information is averaged over several ill-informed individuals, and it is too noisy (31). A recent model for collective behaviour (15) locates the optimal range for anti-predatory response in 2D between 3 and 5 individuals, to be compared to our 3D value of 6-7.

Why a topological, and not a metric interaction? Animal collective behaviour is staged in a troubled natural environment. Hence, the interaction mechanism shaped by evolution must keep cohesion in the face of strong perturbations, of which predation is the most relevant. We believe that topological interaction is the only mechanism granting such *robust* cohesion, and therefore higher biological fitness. A metric interaction is inadequate to cope with this problem: whenever the inter-individual distance became larger than the metric range, interaction would vanish, cohesion would be lost, and stragglers would 'evaporate' from the aggregation. A topological interaction, on the opposite, is very robust, since its strength is the same at different densities. By interacting within a fixed number of individuals, rather than meters, the aggregation can be either dense or sparse, change shape, fluctuate and even split, yet maintaining the same degree of cohesion.

To support this hypothesis we analyze topological vs. metric interaction in the context of one of the simplest two-dimensional flocking models, the Self-Propelled Particles (SPP) model of (17) (see caption of Fig.4 for the equations defining the model). The standard SPP model is strictly metric: each individual interacts with all neighbours within a fixed metric range $r_c$. The model, however, can easily be modified to become topological: each individual interacts with a fixed number of neighbours, $n_c$. In absence of external perturbation, both interactions produce cohesive flocks in an appropriate range of parameters. However, it is not simply cohesion we are after, but *robust* cohesion. We therefore expose a cohesive flock to an external perturbation that mimics the attack of a predator (see Fig.4). A possible outcome of the attack is to break the original flock into many components (Fig.4a). Most of these *M* components consist of isolated individuals, or small groups, which are of course very vulnerable to predation. A robust interaction must preserve cohesion under attack, and thus keep the number of components *M* as low as possible. *M*=1 indicates that the original flock resisted the attack as a whole, and it corresponds to maximum cohesion. Cohesion of the aggregation is therefore higher the lower the number of components *M* after the attack. We perform the numerical experiment a large number of times, with different initial conditions, and compute the probability of having *M* flocks after the attack.

Metric flocks very often break into more than one component, with a maximum probability at $M=5$ (Fig.4b,d). This means that the average resilience of a metric flock is extremely poor: many isolated birds and small groups are forced out of the main flock by the predator's attack. Cohesion in topological flocks, on the other hand, is far superior (Fig.4c,e). The highest probability is at $M=1$, namely the most probable outcome of the attack is that the original flock *does not* break up. Moreover the probability decays very rapidly to zero. Flocks ruled by a topological interaction are therefore much more stable under perturbations than metric ones. We repeated the same experiment using an inter-individual interaction that decays as the inverse of the distance, either metric ($1/r$) or topological ($1/n$). Results were exactly the same (Fig.S6 in SM). This proves a very important point: the nature of the interaction (metric vs. topological) is much more relevant than the specific way it depends on the distance (flat vs. decaying). We expect the difference between topological and metric interaction to be even more striking in 3D, where achieving cohesion is more difficult due to the larger number of individual degrees of freedom.

In conclusion, we presented large-scale 3D empirical data on a paradigmatic instance of collective animal behaviour, namely starlings' aerial display over the roost. Our results show that the inter-individual interaction depends on the topological distance, not the metric distance, at variance with most current models and theories. We suggest that models should be reconsidered in the light of this result. We also argued that a topological interaction is necessary to sustain strong density fluctuations and to maintain cohesion under perturbation, most conspicuously predation. Our numerical simulations support this idea in a compelling way. Given the strong adaptive advantage of cohesion for all animal aggregations, it seems likely that topological interaction is a fundamental ingredient also of other instances of collective animal behaviour. New empirical observations of different systems are necessary to confirm this idea.

**Materials and Methods**

**Location and materials.** Images were taken in Rome, from the terrace of Palazzo Massimo, Museo Nazionale Romano, facing the roost trees situated in the square in front of Termini railway station. The apparatus was located 30m above ground level. Wind speed never exceeded $12 ms^{-1}$. Average distance of birds was 100m. We used Canon EOS 1D Mark II digital cameras (3504×2336 pixels), mounting Canon 35-mm lenses. Focal length, optical centre, radial and tangential distortion were calibrated by Menci-Software s.r.l.. Aperture was between f2.0 and f4.0; shutter speed between 1/1000s and 1/250s; ISO between 100 and 800. We used Manfrotto-400 micrometric heads and Manfrotto-475 tripods. Cameras' tilt-up (between 35% and 40%) was measured by means of a Suunto clinometer.

**Experimental technique.** We used stereo photography (32,33). The distance between stereo cameras (baseline) was $d=25m$. A third trifocal camera was placed at 2.5m from the right stereo camera (Fig. S5). The error $\delta z$ on the *relative* distance of two nearby targets located at distance $z$ from the cameras is dominated by the error $\delta s$ in the determination of the images positions. For parallel focal planes we have,

$$\delta z = \frac{z^2}{\Omega d} \delta s ,$$

where $\Omega$=4335 is the focal length (measured in pixels) of our cameras, and nominally $\delta s$=1pixel. For birds at a distance $z$=100m we thus get a nominal error $\delta z$=0.09m. The error $\Delta z$ on the *absolute* distance of a target at distance $z$ is dominated by the error $\delta\alpha$ on the convergence angle between the focal planes of the stereo cameras,

$$\Delta z = \frac{z^2}{d} \delta\alpha .$$

Each stereo camera was mounted on and aligned to a 680mm-long aluminium bar. A thin ($\varnothing$0.25mm) line run along the bars and connected the stereo pair (Fig. S5). The nominal alignment error was thus $\delta\alpha$=0.25/680=3.7×10$^{-4}$ radiant, thus giving a nominal error $\Delta z$=0.14m (targets at 100m). Regular tests, performed with laser-metered targets gave: $\delta s$<0.4pixel $\Rightarrow$ error on relative distance $\delta z$<0.04m; $\delta\alpha$<2.3×10$^{-3}$radiants $\Rightarrow$ error on absolute distance $\Delta z$<0.92m (for $z$=100m). Stereo cameras were slightly convergent (0.22 radians), to have maximal overlap of the fields of view (Fig. S5).

**Temporization.** The shutter release cables were connected to a timer, which fired the cameras simultaneously at 5 frames-per-second (fps). Synch error was measured in lab and resulted smaller than 10ms. At rates faster than 5 fps the release time lag of the cameras became erratic. To increase shooting rate two interlaced cameras were mounted on each bar (Fig.S5). Thus, our apparatus was shooting at 10fps. Buffers of the cameras filled up after 40 photographs. Therefore each flocking event lasts at most 8 seconds.

**Stereo Matching.** After subtraction of the background, a segmentation algorithm locates birds' positions on the photographs (34). To perform the 3D reconstruction, each bird's image on the left photo must be matched to its corresponding image on the right photo (Fig.1). For large and compact sets of featureless points this problem becomes extremely severe, and it has been the main bottleneck in the 3D reconstruction of animal aggregations. Our matching procedure involves 3 cameras, A, B and C (Fig. S5) and has four steps: first, a newly developed algorithm, exploiting pattern-recognition and epipolar invariance (33), matches approximately 20% of the birds on the stereo pair (A-C). Secondly, the same algorithm matches approximately 90% of the birds on the two nearby cameras (A-B). Third, these matches are used to calculate the nonlinear trifocal tensor (35). Fourth, by means of the trifocal tensor and of an optimization assignment algorithm (36), the (A-B) matches are transferred to the pair (A-C). In the cases we analyzed, we match on average 88% of the birds, and never less than 80%. Tests with synthetic data (distance and density as in the biological case) gave less than 5% of mismatches (outliers). A framework software developed by Nergal s.r.l. coordinated the various parts of the code. Our algorithms use some routines from the LTI-lib (http://ltilib.sourceforge.net) and CGAL (http://www.cgal.org) libraries.

**Events selection.** We collected approximately 500 flocking events. The vast majority of these events had to be discarded either because they were not included in the field of view of all 6 cameras, or because they were too far (our photographic resolution requires that flock are closer than 250m). Moreover, many events were recorded in too severe light conditions. These criteria are non-biased, and do not affect the biological features of the flocks, so that the ~50 events remaining are a fair sampling of roost's flocks. We then narrowed our target of investigation by selecting 10 out of these 50 events. We chose flocks with sharp borders, strong spatial cohesion and a large number of birds (>400). We also had to discard flocks that were too dense, since our matching algorithm put a limit to the maximum density (the constraint on density will be relaxed as newer versions of our software are developed). Therefore, the final 10 analyzed events are representative of the typical *cohesive* flocks over the roost. We finally checked on synthetic data that the reconstruction software does not introduce any significant bias in the flock's shape and structure. All the reconstructions belonging to a single event are statistically homogeneous, and were thus used to build statistics for that particular flock.

**The problem of the border.** Flocks are not necessarily convex. Therefore the so-called *convex hull* is not a suitable tool to define their border. In order to define the border we used the $\alpha$-*shape* algorithm (37): basically one excavates the set of points with spheres of radius $\alpha$. For $\alpha=\infty$ the border coincides with the convex hull, whereas for finite $\alpha$ concavities of size $\alpha$ are detected. Once the border's points are defined, one has to take care of the bias introduced by them. All methods to eliminate such bias basically require excluding the border points from the analysis, and for this reason having large aggregations is essential. We used the Hanish method (38). According to this method, when computing a certain average quantity at a given scale $r$, only the points having a distance from the border greater than $r$ are considered. We checked all our tools by using them in test cases (as the Poisson and the hard-spheres distribution) where the analytic results were known. In these test cases one gets wrong results if the border's bias is not eliminated.

**Acknowledgements** We thank E. Alleva for several crucial discussions and for the invaluable help he gave us on all ornithological issues, together with C. Carere and D. Santucci. We are indebted to C. Agrillo, E. Branchini, F. Cecconi, M. Cencini, I.D. Couzin, A. Gabrielli, R. James, T. Joerg, J. Krause, N. Leonard, B. Marcos, E. Marinari, M. Montuori, R. Sarno, F. Sylos-Labini, and to all members of the STARFLAG project for discussions. We thank S. Cabasino, F. Ceccaroni, A. Cimarelli, M. Fiani, L. Menci, C. Piscitelli, R. Santagati, F. Stefanini for technical help, and R. Paris and M. Petrecca for granting the indispensable access to Palazzo Massimo, Museo Nazionale Romano. We also thank P. Corezzola, G. Loffredo, M. Marchetti and B. Pernati for administrative assistance. We finally thank G.A. Cavagna for carefully reading the manuscript and for the advice. V.L. acknowledges the kind hospitality of the Department of Physics of the University of Rome La Sapienza, where part of this study was performed. This work was financed by a grant from the European Commission to the STARFLAG project, within the NEST-PATHFINDER initiative under Framework Programme 6.

**Author Contributions** N.C., A.C., G.P. designed the experiment; M.B., E.C., N.C. designed and built the electronics; M.B., R.C., A.P., V.Z. performed the experiment under A.C. supervision; A.O. was responsible for computer vision and trifocal method, and with A.C., G.P., M.V. developed the segmentation and reconstruction algorithms; A.C. and I.G. designed the pattern-recognition algorithm, M.V. improved on it and wrote the code; A.C., I.G., G.P. coordinated the data analysis; A.C., I.G., A.O., A.P. performed the data analysis. A.C., I.G., V.L. designed the simulation; V.L. performed the simulation. The paper was written by A.C., with the participation of I.G. and G.P. The project was coordinated by A.C.

# Figures

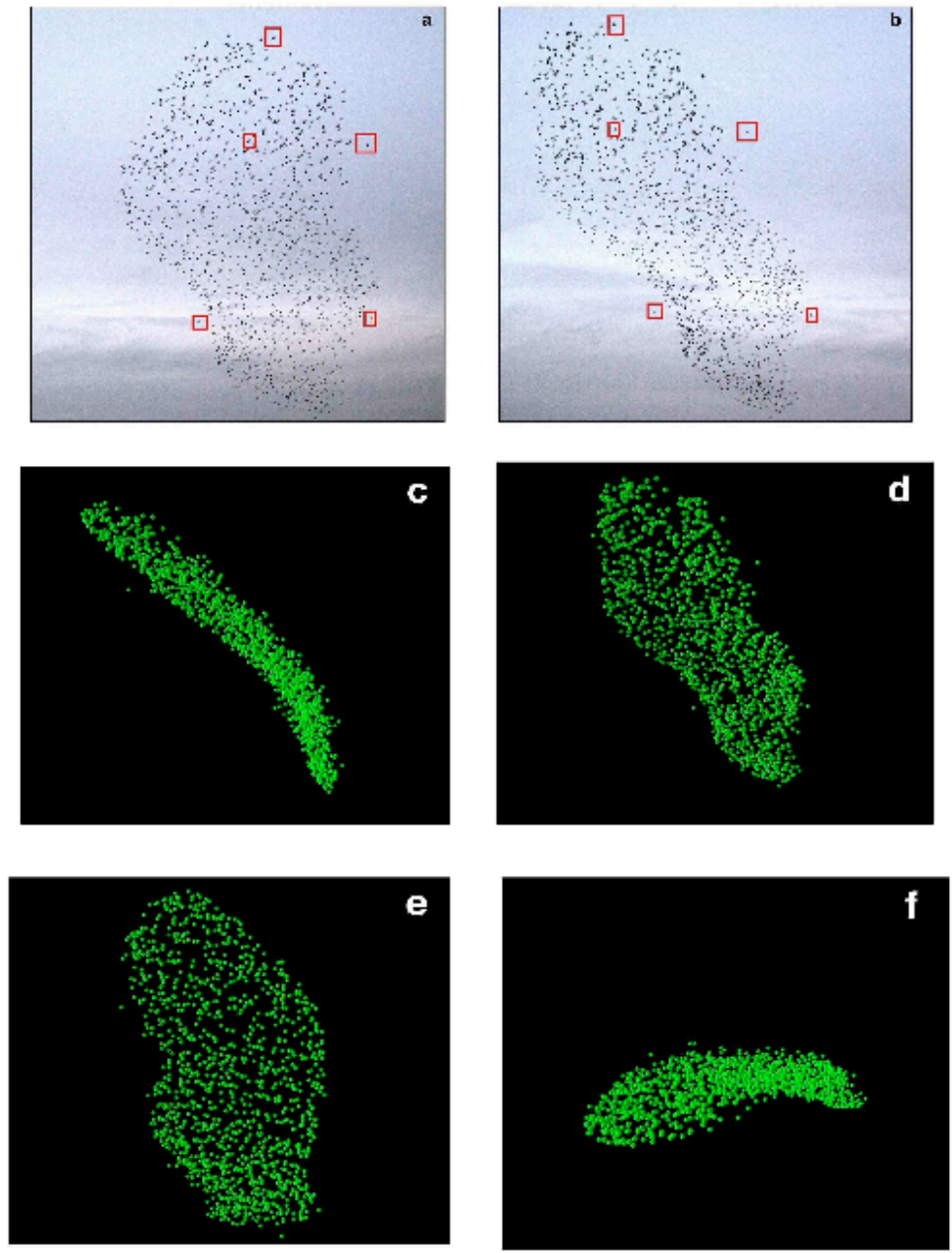

**Fig. 1.** A typical analyzed flock. This aggregation consists of 1246 starlings, flying at approximately 70m from the cameras at about 11ms$^{-1}$ (flock 28-10 in table S1). **a,b,** Left and right photographs of the stereo pair, taken at the same instant of time, but 25 meters apart. To perform the 3D reconstruction, each bird's image on the left photo must be matched to its corresponding image on the right photo. Five matched pairs of birds are visualized by the red squares. **c,d,e,f**, 3D reconstruction of the flock under 4 different points of view. Panel **d** shows the reconstructed flock under the same perspective as the right photograph (**b**).

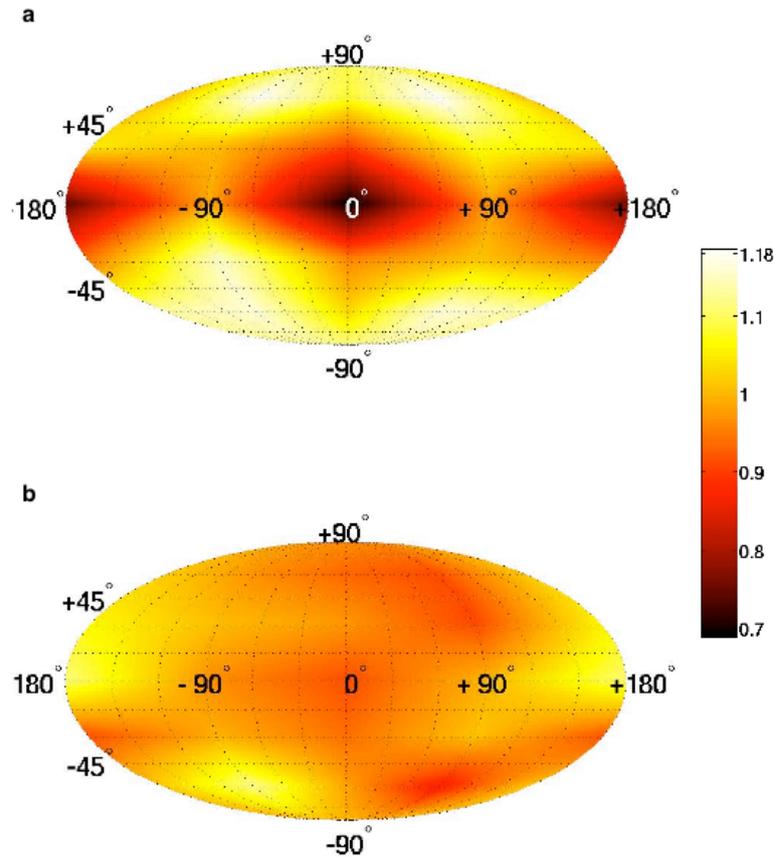

Figure 2

**Fig. 2a.** The angular density of nearest neighbours shows a strongly anisotropic structure of individuals. This density is proportional to the probability of finding the nearest neighbour of a bird at certain angles ($\phi$, $\alpha$) around it. The map is obtained as follows. For each bird *i* we define the unit vector $\mathbf{u_i}$ in the direction of its nearest neighbour. We then place the vectors corresponding to all nearest neighbours at the same origin and plot their density on the unitary sphere. The spherical density is projected on the plane by means of an equal-area Mollweide projection. We have normalized by the isotropic case, so that the density is uniform and equal to 1 for a non-interacting aggregation of individuals. The reference frame is chosen in the following way. The flock's velocity **V** is the first reference direction. Since the velocity **V** and gravity **G** are approximately orthogonal in all the flocks we analyzed (we find on average **V·G**=0.13±0.02 SE), it is useful to consider as a second reference direction $\mathbf{G_\perp}$, the component of gravity perpendicular to the velocity. In the map, the velocity **V** goes through the centre of the map and $\mathbf{G_\perp}$ corresponds to the zenith of the map, whereas the plane P orthogonal to $\mathbf{G_\perp}$ corresponds to the horizon. The latitude, or *elevation*, $\phi \in$ [-90:90] indicates the angle in degrees between $\mathbf{u_i}$ and the horizon plane P. The longitude, or *bearing*, $\alpha \in$ [-180:180] indicates the angle in degrees between the projection of $\mathbf{u_i}$ on the horizon plane P and the velocity **V**. Therefore, the centre of the map ($\phi$=0, $\alpha$=0) corresponds to the front of the bird, whereas the points ($\phi$=0, $\alpha$=+180) and ($\phi$=0, $\alpha$=-180) correspond to the rear of the bird. The density is strongly anisotropic, with a significant lack of nearest neighbours along the velocity. The map is calculated using data from one particular flock (flock 25-11, see table S1 in SM). However, data from the other flocks *all* show the same lack of nearest neighbours along the velocity (Fig. S4 in SM). On the other hand, the location of the maxima, which may suggest a structural correlation with the plane P orthogonal to gravity, is less stable from flock to flock (Fig. S4 in SM). A more detailed analysis of this feature is required before drawing any conclusion.
**2b.** The density the 10$^{th}$-nearest neighbour shows no statistically significant structure and it is compatible with a set of non-interacting points. This indicates that interaction with the 10$^{th}$-nearest neighbour has completely decayed.

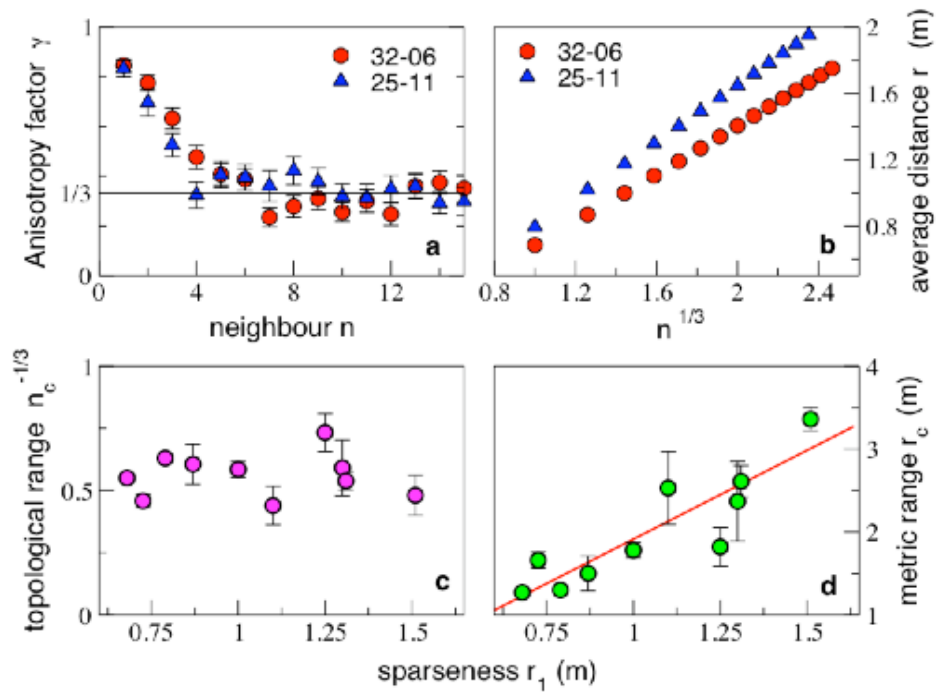

**Fig.3.** Assessing the range of the interaction. Let $\mathbf{u}_i^{(n)}$ be the unit vector pointing in the direction of the $n^{th}$-nearest neighbour of bird $i$. We define the projection matrix $\mathbf{M}^{(n)}$ as,

$$\mathbf{M}^{(n)} = \frac{1}{N}\sum_{i=1}^{N} \mathbf{u}_i^{(n)} \otimes \mathbf{u}_i^{(n)} \quad , \quad \mathbf{w} \otimes \mathbf{v} = w^\alpha v^\beta \quad , \quad \alpha,\beta = x,y,z$$

where $N$ is the number of birds in the flock. The unitary eigenvector $\mathbf{W}^{(n)}$ relative to the *smallest* eigenvalue of $\mathbf{M}^{(n)}$ coincides with the direction of minimal density of the vectors $\mathbf{u}_i^{(n)}$, i.e. the direction of minimal crowding of the $n^{th}$-nearest neighbour. To measure the degree of anisotropy in the spatial distribution of the $n^{th}$-nearest neighbour we use the function $\gamma(n)=(\mathbf{W}^{(n)} \cdot \mathbf{V})^2$, where $\mathbf{V}$ is the velocity. The value of $\gamma$ for an isotropic, non-interacting distribution of points is 1/3. A value of $\gamma(n)$ larger than 1/3 indicates that there is a lower probability to find the $n^{th}$-nearest neighbour along the direction of motion, and thus indicates anisotropy. **a.** The function $\gamma(n)$ is plotted for two different flocks (32-06 and 25-11); error bars represent the standard error. For both flocks the structure becomes approximately isotropic between the $6^{th}$ and the $7^{th}$-nearest neighbour. The topological range $n_c$ is defined as the point on the abscissa where a linear fit of $\gamma(n)$ in the decreasing interval intersects the value 1/3. **b,** The average distance $r_n$ of the $n$-th neighbour is plotted against $n^{1/3}$, for the same flocks as in **a** (error bars are smaller than symbols size). The linear relation between $r_n$ and $n^{1/3}$ is very sharp in all flocks. The slope of these curves is proportional to the sparseness $r_1$ of the flock, i.e. the average nearest neighbours distance. **c,** Topological range $n_c$ (to the power $-1/3$) vs. the sparseness $r_1$ of each flock. No significant correlation is present (Pearson's correlation test: $n=10$, $R^2=0.00021$, $P=0.97$). **d,** Metric range (in meters) $r_c$ vs. sparseness $r_1$. A clear linear correlation is present in this case ($n=10$, $R^2=0.78$, $P<0.00072$). This proves that the natural scale of the interaction is the topological distance, and not the metric one. The topological range is (on average) between 6 and 7 neighbours.

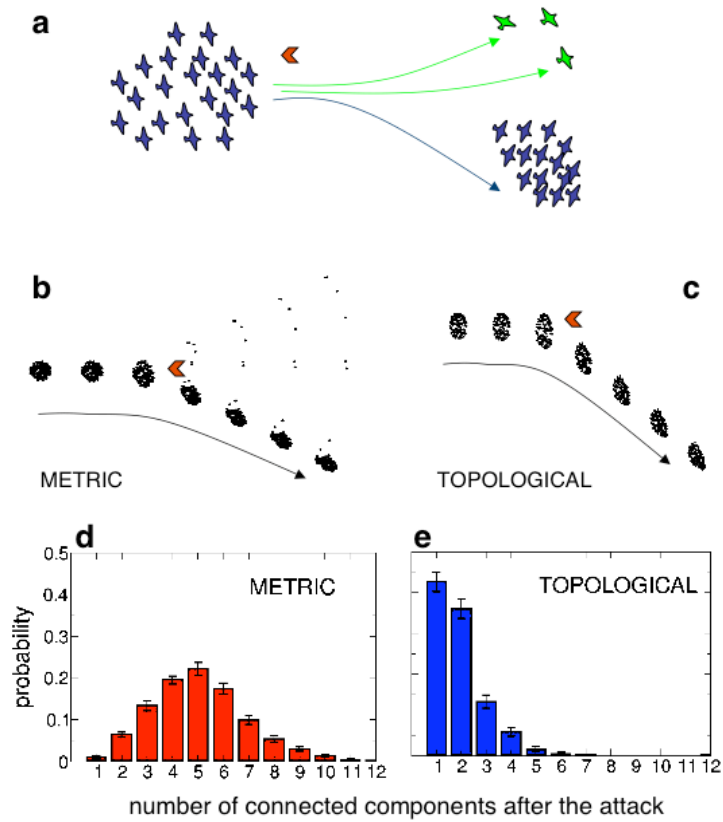

**Fig. 4**. Numerical simulations: metric vs. topological interaction under predator's attack. We used the SPP model of (17): a set of 'birds' in 2 dimensions moves synchronously and interact by aligning with their neighbours. Each bird $i$ is characterized by its position $r_i$ and velocity $v_i$. The dynamics is defined by $r_i(t+1) = r_i(t) + v_i(t+1)$. The velocities $v_i$ have constant modulus, whereas their heading $\theta_I$ obey the equation, $\theta_i(t+1) = [\theta_i(t) + \Sigma_j \theta_j(t)]/(N_i+1)$, see also (6). The sum runs over the $N_i$ neighbours interacting with bird $i$. In the *metric* version of the model one considers all neighbours within a fixed metric range $r_c$ around bird $i$, whereas in the *topological* case only the first $n_c$ neighbours are considered, irrespective of their metric distances: $N_i = n_c$ for all $i$. The flock and the predator are put in relative motion one against the other, with a vertical offset $d$ between the predator and the centre of mass of the flock. The predator exerts a repulsive force on each bird, which decays with the bird-predator distance as $1/r$ and gives a contribution $F_0 [y_i \cos(\theta_I) - x_i \sin(\theta_I)] / r_i^2$ to the equation for the heading $\theta_I$. The parameter $F_0$ tunes the strength of this force relative to the inter-individual one. **a** Sketch of the experiment: an initially cohesive and polarized flock moves towards the predator (orange arrow) and interacts with it. **b,c** Typical flocks' trajectories. In the metric case many birds are pushed out of the flock, and remain isolated. Conversely, in the topological case the flock maintains cohesion, and no stragglers arise. **d,e** Probability that the flock breaks into $M$ connected components (CC) after the attack; a CC is defined as a set of birds that are within a distance $3r_c$ from at least one bird. The value $M=1$ indicates that the original flock resisted the attack as a whole, i.e. maximum cohesion. The probability is dramatically different in the two cases. Moreover, in the metric case stragglers are the 43% of the CC, whereas in the topological case they are just the 5%. In a second simulation, flocks are sent against an obstacle. In order to avoid it, the velocity of each bird is *randomly* reassigned whenever it gets too close to the obstacle. The probability of $M$ is very similar to the predator setup; metric stragglers are 24% of the CC; topological stragglers are 0.7%. These results prove that topological flocks are significantly more resilient than metric flocks to perturbations. Parameters of the simulation are: N=200 particles; T=2000 time steps; number of different initial conditions $N_{in}$= 5000 (metric case), 2000 (topological case); $r_c$=0.15 (metric case); $n_c$= 3 (topological case); $|v_i|$= 0.25s$^{-1}$; $d$ =0.9; $F_0$=0.05. Initial birds are confined in a region of size $R$=1, and have aligned velocities. Boundaries are open. We checked that the results do not change qualitatively in an ample and stable range of parameters.

# SUPPLEMENTARY MATERIAL

### What is the origin of the anisotropy?

A similar spatial anisotropy as the one we find for starling was reported in fish schools (39), and it seems therefore a typical feature of collective behaviour in both birds and fishes. What is the origin of this anisotropy?

Numerical models with isotropic interaction break the directional symmetry, giving a non-zero velocity of the aggregation, but fail to reproduce the structural anisotropy (40-41). This suggests that the anisotropy is not simply an effect of the existence of a preferential direction (the velocity), but is rather an explicit consequence of the anisotropic character of the interaction itself. Vision is a natural candidate, given its anisotropic nature in both birds and fishes. In particular, starlings have lateral visual axes and a blind rear sector (42), and this fact is likely to be related to the lack of nearest neighbours in the front-rear direction. Indeed, several studies interpreted the anisotropic flight formations in birds as the result of the optical characteristics of the birds' eye (43,44,45). To investigate further this hypothesis, it would be very important to have an argument that connects in a quantitative way the physiological field of view of the birds to the actual position of the nearest neighbours. Unfortunately, there is no such model to-date. A distinct idea is that the mutual position chosen by the animals is the one that maximizes the sensitivity to changes of heading and speed of their neighbours (46). According to this hypothesis, even though vision is the main mechanism of interaction, optimization determines the anisotropy of neighbours, and not eye's structure.

Radically different is the claim that anisotropic structures both in bird and in fish aggregations save energy thanks to aerodynamic (or hydrodynamic) advantages (AA) (47,48,49). In particular, this hypothesis has been often invoked to explain V-formations in migrating birds. It may be plausible to advocate an energy-saving principle for migrating birds, but it seems much less so for flocking starlings, when birds appear to do everything but economize on energy. In fact, the energy-saving principle has been challenged both for birds (50) and fishes (51). Our discovery that flocking is ruled by a topological interaction rules out the AA hypothesis, at least for starlings in evolution over the roost. Indeed, topological interaction and AA are completely incompatible: in order for AA to work, animals must keep roughly constant mutual positions in space, because aerodynamics is ruled by *metric* scales. In the topological case however, the anisotropy is the same for dense as for sparse flocks, and the mutual metric positions of birds change drastically with the density. We cannot exclude that for migrating birds things are different. In that case, the aggregations may keep a constant density in time, making the AA thus possible. Only empirical observations can clarify this question.

### Topological range or renormalized metric range?

Our experimental results show that the interaction among the individuals depends on the topological distance, rather than the metric one. This means that, given two flocks with different density, the topological range is the same in the two flocks (say 6 birds), whereas the metric range is large in the sparse flock and smaller in the denser flock. Formally, we can introduce a *renormalized* metric range, by dividing the normal metric range by the average nearest neighbours' distance, i.e. the sparseness, $r_c' = r_c/r_1$. Of course, this quantity is nothing else that the topological range to 1/3, $r_c'=n_c^{1/3}$. The renormalized metric range is therefore the same in the two flocks, as it does not depend on sparseness anymore.

From the mathematical point of view, introducing a topological range, or a renormalized metric range is equivalent. There are, however, many reasons, both technical and biological, to prefer the topological vs. metric description, rather than the renormalized metric vs. purely metric one. To implement the topological interaction in a simulation, one simply makes each bird interact with all its $n_c$ nearest neighbours, which is technically easy to do. On the other hand, one could compute the metric distance of the neighbours, compute the average nearest neighbours' distance in the whole flock, divide the two, and make the bird interact with all birds within a certain renormalized metric distance. This is much more complicated to do. Secondly, a topological range is more suitable to spell out the cognitive implications of our result. The topological range of the interaction (6-7 neighbours) is the same as the threshold of numerosity discrimination found in experiments on birds (7 objects). It would harder to appreciate this result by saying that the renormalized metric range of interaction is constant, and equal to 1.5 (for example). The renormalized range would be dimensionless, and it would very poorly connect with experiments on the cognitive capabilities of birds.

**The connection between structure and interaction**

Several numerical models of collective behaviour are ruled by completely isotropic interactions and, consequently, do not display any anisotropy in the spatial structure of individuals. Yet, individuals are strongly interacting. One may then argue that there is no link, in general, between structure and interaction. However, this is not true. The fact that interaction in real flocks gives rise to anisotropy in the structure is just a fortunate condition that allows us to trace the interaction in a very straightforward way, and find its range. Had the interaction *not* been anisotropic, it would just have been harder to trace the interaction from the structure, but still quite possible; for example, by carefully analyzing the so-called structure factor as it is normally done in liquid theory. So, in order to read the interaction from the structure, it is important to find the right tracer. For flocking birds, we found that the anisotropy is a very good tracer.

A second possible objection is that birds, in addition to the (probably) visual anisotropic interaction that we identify, may also be using a different kind of interaction, which however leaves no trace on their structure. In this case, of course, we would not be able to say anything about this second 'invisible' interaction, and our analysis would be partial. To this objection one can reply first of all that an interaction that has no effects on the structure of individuals seems significantly less relevant than one that does. Secondly, even though we cannot exclude such case, it seems more reasonable to keep the number of possible ingredients to a minimum, and not to invoke the existence of forces of which we find no effects in the empirical data.

Finally, we must remark that in this work we only studied the *static* structure of individuals, leaving to a future work the analysis of the individual trajectories, with all *dynamical* correlations. Dynamics is of course another very important effect of the interaction, so it will be very interesting to compare the results coming from the dynamical structure (range, interaction's decay, etc) with the present static results.

## Supplementary Figures

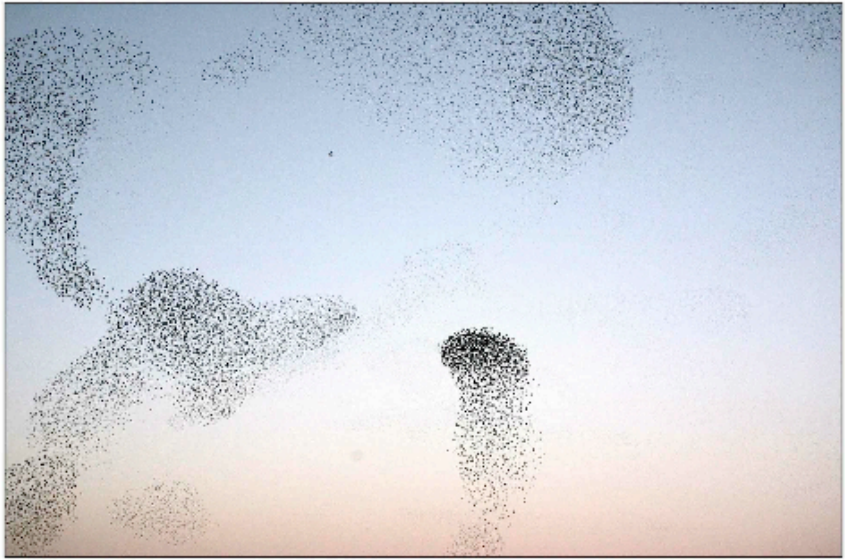

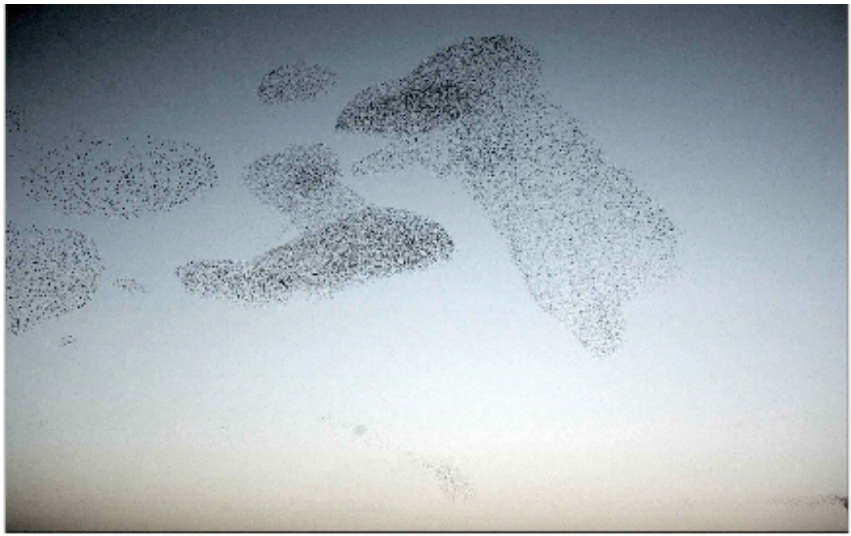

### Figures S1-S2. Aerial display

Typical examples of starling aerial display. Various flocks, ranging from several hundreds to thousands birds, wheel and turn above the roosting site.

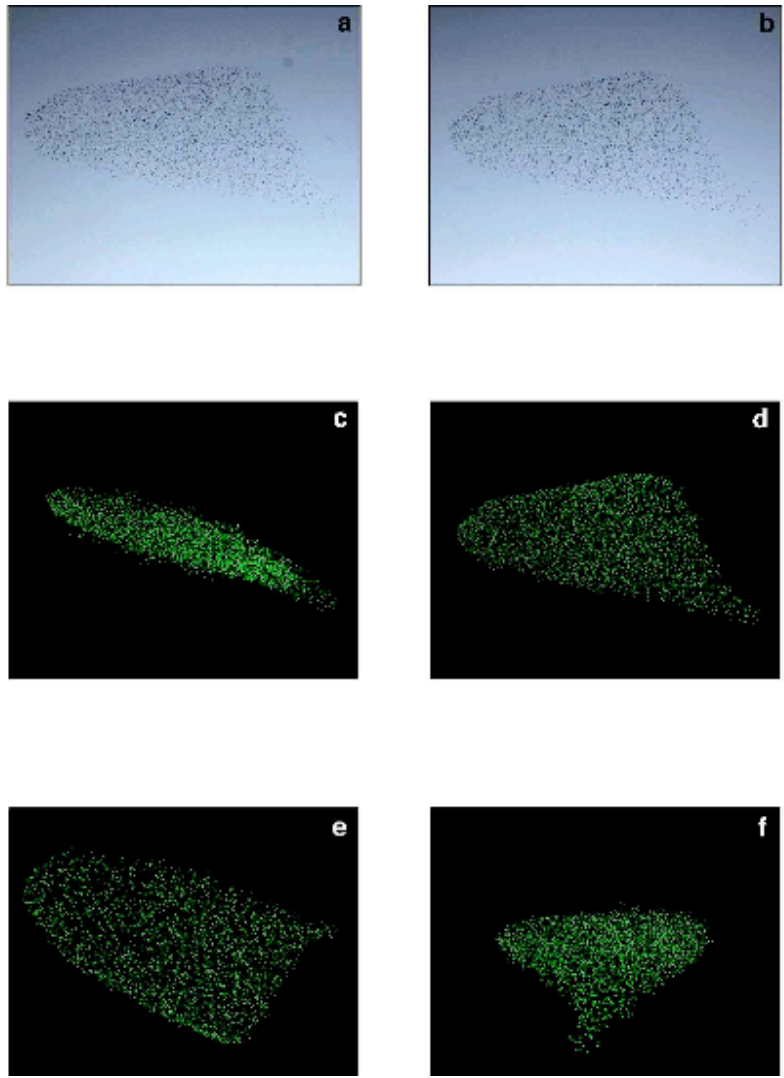

**Figure S3. Flock image and 3D reconstruction**

This flock (16-05) consists of 2631 starlings, flying at approximately 240 m from the cameras. The cameras tilt-up was 40%. **a,b**, Left and right photographs of the stereo pair, taken at the same instant of time, but 25 meters apart. **c,d,e,f**, 3D reconstruction of the flock in the reference frame of the right camera, under 4 different points of view. Panel **d** shows the reconstructed flock in the same perspective as the right photograph (**b**).

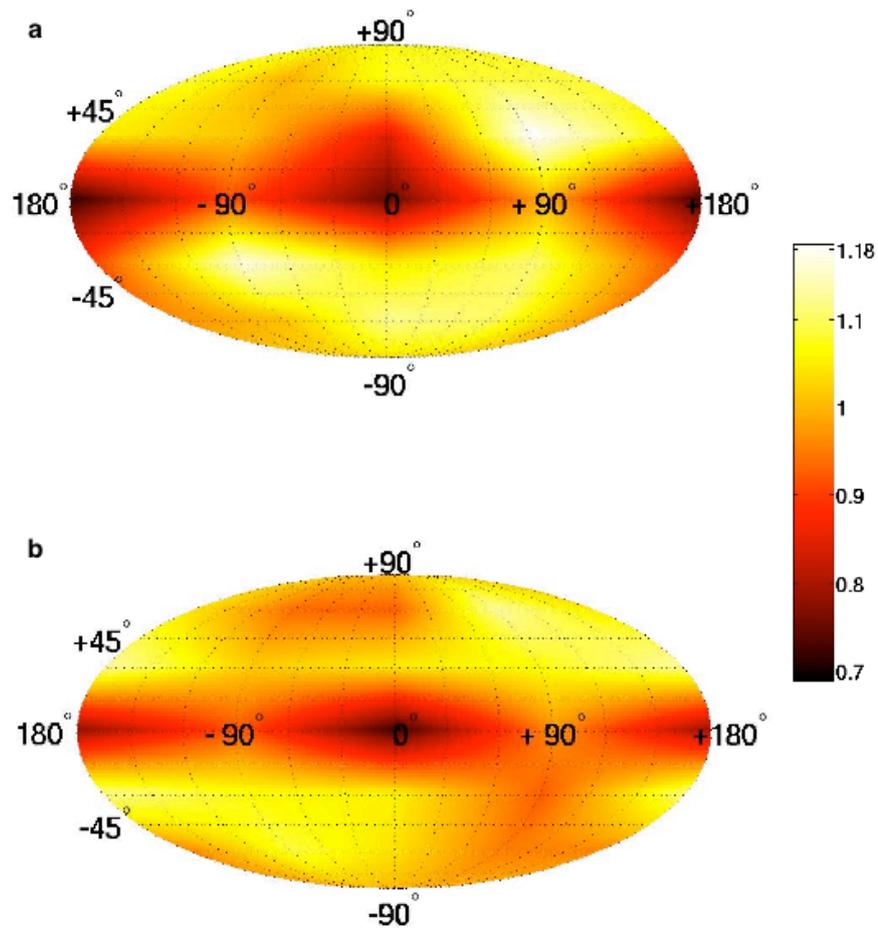

**Figure S4. Angular distribution map of nearest neighbours.**

This figure shows the angular distribution map of nearest neighbours for two flocks (panel **a** and **b**). The map is computed as described in Fig.2a. In both flocks there is a strong lack of neighbours along the velocity. The location of the maxima is different in the two cases. With reference to Table S1, the maps in this figure correspond to events 21-06 (panel **a**) and 29-03 (panel **b**).

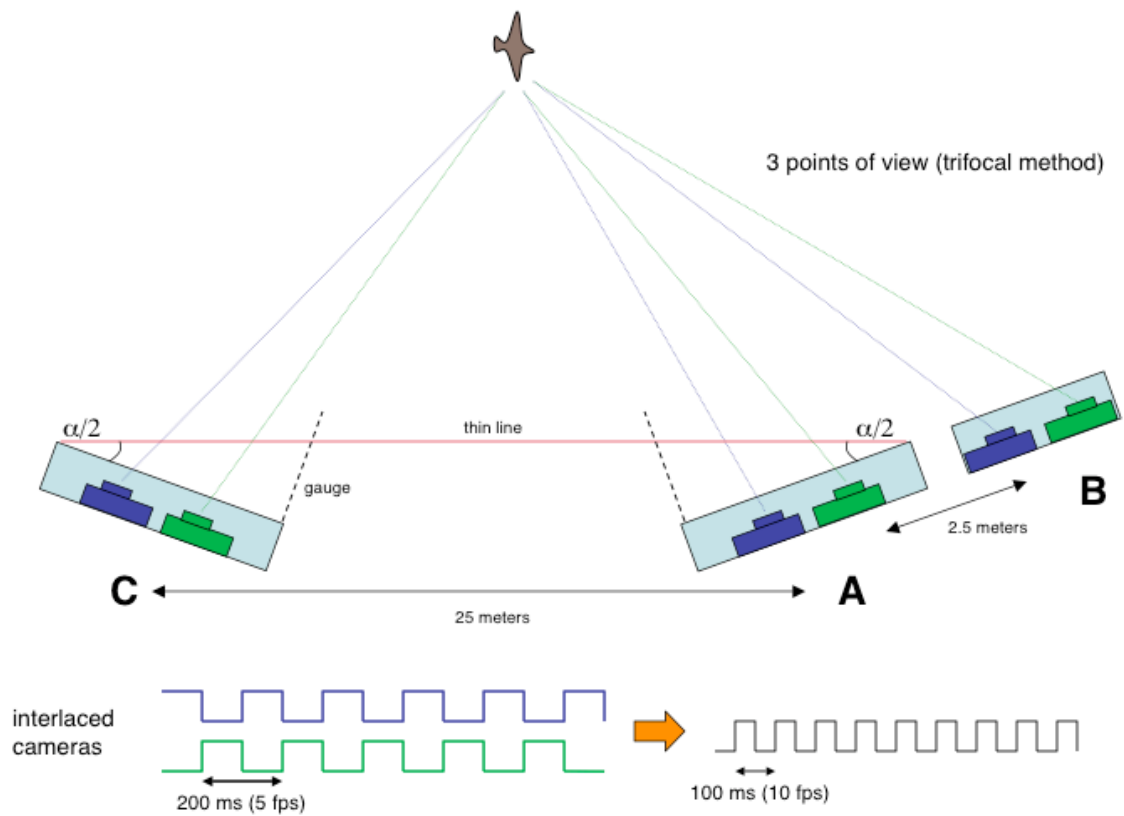

**Figure S5. Experimental setup**

Schematic illustration of the experimental setup. There are three observation points where the cameras are located: two of them are 25m apart (stereo cameras) and the third one is 2.5m further on the right (trifocal camera). In each observation point a pair of interlaced cameras (green and blue in the figure) is mounted on and aligned to an aluminium bar. For stereo cameras the bar is 680mm long and is equipped with a rigid graduated gauge. Bars have a convergence angle $\alpha$ of 0.22 radiants to optimize the common field of view of the cameras and a tilt-up which may vary from 35% to 40%. A thin line (⌀0.25mm) connects the stereo bars passing close enough to the gauge to avoid parallax effects. In this way the nominal error on the convergence angle is $3.7 \times 10^{-4}$ radiants. The left (blue) cameras on the bars shoot simultaneously with a period of 200ms (5fps). The same is true for the right cameras (green). Left and right cameras are interlaced with a shift of half a period such that the global shooting rate is 10 fps.

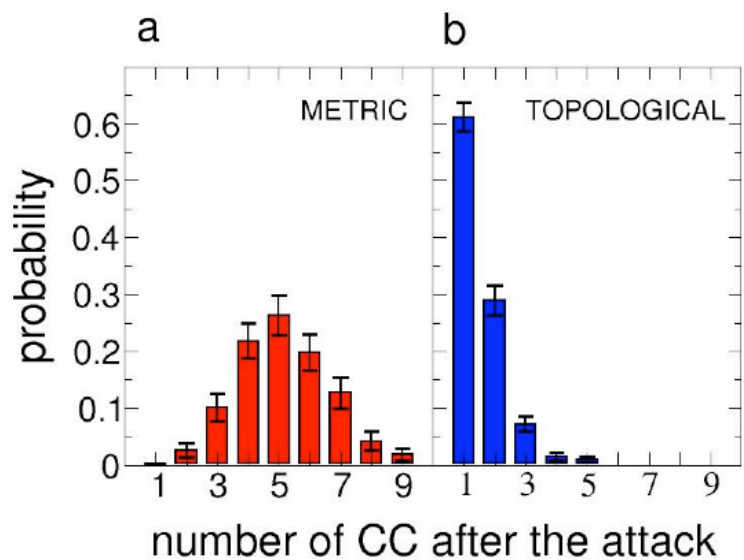

**Figure S6. Simulations with interaction decaying with the distance**

Numerical simulations: probability of the number of connected components (CC) after the predator's attack, in the metric vs. topological case, with an interaction among individuals that decays with the distance. The setup of the simulation is the same as Fig.4, but the way the directions of the velocities are updated is different. In Fig.4 all neighbours, both in the metric and topological case, were weighted the same in the update of the headings $\theta_i$. Here, however, we weight each bird with a term $1/(1+r_i)$ in the metric case, and $1/(1+n_i)$ in the topological one. So, birds that are more distant from the reference one, interacts more weakly with it, but distance can either be topological or metric. As in the flat, un-weighted case, the probability is dramatically different in the two cases, with topological flocks much more cohesive than metric ones under perturbation. Therefore the big difference is between the metric and the topological case, and not between the flat and the decaying case. This proves that to keep cohesion under perturbation, the definition of distance (metric vs. topological) is definitely more relevant than the specific dependence on it (flat vs. decaying).

## Table S1. Global quantitative properties

Flocking events are labelled according to session number and position within the session. Each quantity is averaged over the different shots of the event. Events are ordered by increasing values of the average nearest neighbour distance, $r_1$, which we call *sparseness* throughout our study. The density $\rho$ (the ratio of the number of birds to the flock's volume) is proportional to $r_1^{-3}$ (Pearson's correlation test: $n=10$, $R^2=0.8$, $P=0.0004$). The value of $r_1$ varies significantly, ranging from 0.7m to 1.5m, in the analyzed events. Average body length and wingspan of starlings are respectively BL=0.2m and WS=0.4m, and thus in the densest flocks $r_1 \sim 3.5$BL and $r_1 \sim 1.7$WS. Velocity refers to the centre of mass (the average position of all birds in the flock). The scalar product between gravity and velocity is always quite small: velocity is thus approximately perpendicular to gravity.

| Event | Number of birds | Sparseness $r_1$ (m) | Velocity (ms$^{-1}$) | \|V*G\| |
|---|---|---|---|---|
| 32-06 | 781 | 0.68 | 9.6 | 0.06 |
| 28-10 | 1246 | 0.73 | 11.1 | 0.09 |
| 25-11 | 1168 | 0.79 | 8.8 | 0.12 |
| 25-10 | 834 | 0.87 | 12.0 | 0.18 |
| 21-06 | 617 | 1.00 | 11.2 | 0.09 |
| 29-03 | 448 | 1.09 | 10.1 | 0.27 |
| 25-08 | 1360 | 1.25 | 11.9 | 0.14 |
| 17-06 | 534 | 1.30 | 9.1 | 0.09 |
| 16-05 | 2631 | 1.31 | 15.2 | 0.19 |
| 31-01 | 1856 | 1.51 | 6.9 | 0.09 |